\documentclass[amsmath,amssymb,onecolumn,superscriptaddress]{revtex4}
\usepackage{graphicx}%
\usepackage{dcolumn}%
\def\be{\begin{equation}}
\def\ee{\end{equation}}
\def\beq{\begin{eqnarray}}
\def\eeq{\end{eqnarray}}

\usepackage{bm}% bold math
\usepackage{graphicx, psfrag}
\usepackage[colorlinks=true, citecolor=blue, urlcolor = blue, linkcolor= red, bookmarks=true]{hyperref}
\usepackage{float}
\usepackage{amssymb}
\usepackage{amsmath}
\usepackage{hyperref}
\usepackage{subfigure}
\usepackage{pgfplots}
\usepackage{epstopdf}
\usepackage{booktabs}
\usepackage{orcidlink}
\usepackage{xcolor}

\begin{document}
\title{Stability of Relativistic Spheres Powered by Energy Profile in Einstein Gauss-Bonnet Gravity}

\author{Z. Yousaf}
\email[Email: ]{zeeshan.math@pu.edu.pk}
\affiliation{Institute of Mathematics, University of the Punjab, Lahore-54590, Pakistan}

\author{Kazuharu Bamba}
\email[Email: ]{bamba@sss.fukushima-u.ac.jp}
\altaffiliation{Corresponding author}
\affiliation{Faculty of Symbiotic Systems Science,
Fukushima University, Fukushima 960-1296, Japan}

\author{Mansoor Alshehri}
\email[Email: ]{mhalshehri@ksu.edu.sa}
\affiliation{Department of Mathematics, College of Science,\\
King Saud University, P.O.Box 2455 Riyadh 11451, Saudi Arabia.}

\author{A. Farhat}
\email[Email: ]{aqsaalirajpoot10@gmail.com}
\affiliation{Institute of Mathematics, University of the Punjab, Lahore-54590, Pakistan}


\begin{abstract}
In this paper, we study the dynamical irregularity of the locally anisotropic spherical fluids in the context of Einstein-Gauss-Bonnet theory. We aim to describe the causes of energy-density irregularity of self-gravitating fluids and explain how those causes evolved from a homogeneous distribution at first. After computing field equations, we formulate two independent components of evolution equations. This expression involves the Weyl tensor and dynamical variables that would lead us to explain the emergence of inhomogeneity patterns. The relevant quantities involved in the irregularities within the initially homogeneous system are analyzed by considering particular non-dissipative and dissipative distribution cases. We find the theoretical irregularity factor consistent with astrophysical observations. With this relation, it is explicitly demonstrated that in the presence of Einstein-Gauss-Bonnet gravity terms, the inhomogeneity term decreases its role gradually as the observer moves away from the center towards the boundary surface.

\end{abstract}
\maketitle

\section{Introduction}

Despite being the most widely accepted and effective theory of
gravity, modifying general relativity (GR) may provide
some interesting results. This would lead to understanding some
interesting outcomes \cite{nojiri2011unified,capozziello2011extended,nojiri2017modified,yousaf2022f,yousaf2023quasi}. Padmanabhan \cite{padmanabhan2008dark} discussed the implications of the cosmological constant as a candidate for dark energy and its role in explaining the nature of gravity. Durrer and Maartens \cite{durrer2008dark} expressed concern over the challenges faced when attempting to explain cosmic acceleration in GR. Although there is a wide range of models to explain cosmic accelerating expansion, they are all unable to take into consideration the gravitational characteristics of the vacuum energy. Gravity itself may deviate from GR on large scales in a way that causes acceleration, providing an alternative to dark energy. They concluded that to address this problem, a whole new paradigm may provide explanations. The $f(R)$ theories (where $R$ is the Ricci scalar), attempting to cover as much of the pertinent literature as feasible while presenting the theory's most significant features thoroughly have been reviewed \cite{sotiriou2010f}.

The several uses of
$f(R)$ theories in cosmic history, including spherically symmetric solutions in the strong and weak gravitational backgrounds, dark energy, inflation, and cosmic perturbations, are described in the literature \cite{de2010f}. They offered several observational and experimental strategies to separate such theories from GR. An incredibly extensive overview of current research on modified gravity and its implications for cosmology, and discussed a variety of topics such as higher-order theories, scalar-tensor, etc, is studied in \cite{clifton2012modified}.  The modified theories and cosmology theories could help to understand the finite-time cosmological singularities \cite{de2023finite}. Some other strategies have been put out in the literature to avoid or lessen finite-time singularities in cosmological contexts. Nojiri and his associates \cite{nojiri2008inflation} introduced $f(G)$ (where $G$ is the Gauss-Bonnet scalar) and described the cosmological implications of correction terms in the study early universe. Nojiri \emph{et al.} \cite{nojiri2021ghost} investigated the ghost problem within $F(R,\mathcal{G})$ gravity and suggested that the framework might be perceived as a reconstruction methodology and employed as a means of achieving several interesting cosmic evolutions.

The most generic gravitational theory that yields conserved equations of motion of second order in any D-dimensional space-time is Lovelock theory. The Lagrangian in Lovelock theory \cite{lovelock1971einstein} can be described as a composition of various quantities, where each term represents the Euler density in a spacetime having 2n-dimension, which is expressed as $\mathcal{L}_{n}(2n<D)$. In spacetime, the $\mathcal{L}_{n}$ term becomes a topological entity as the dimension approaches a critical value of $D = 2n$. The application of the GB term for a better understanding of gravity is documented in
\cite{wheeler1986symmetric,cai2002gauss,odintsov2020swampland}. A novel 4D-EGB gravity where the GB term does contribute to dynamics was described very recently by Glavan and Lin \cite{glavan2020einstein}. A particular curvature invariant in a given geometry is described by an expression known as the GB term. Glavan and Lin \cite{glavan2020einstein} hoped to gain a more thorough knowledge of gravity in $4D$ spacetime by adding this component to the dynamical nature of the cosmos. The 4D EGB
gravity is intriguing because it incorporates curvature terms of higher order and can be considered a plausible extension of GR.

It is anticipated that these terms will become significant at extremely high energy, like those observed in black holes or the early stages of the universe. Several intriguing characteristics of the theory, including the absence of tachyons and ghosts, which are troublesome in other theories of gravity, have been demonstrated by theoretical physicists through thorough study. The basic principle of EGB gravity is that, in higher-dimensional spacetimes, the extra-dimensional curvature might reveal new information about the nature of gravity and have a substantial impact on gravitational interaction. Then, they tuned the GB-associated coupling with $D$ dimensions in the equations of motion. They rescaled the coupling related to GB term in D-dimensional spacetime, $(\alpha\rightarrow \frac{\alpha}{D-4})$, before taking the limit $D = 4$ in the field equations to retrieve the role of the local dynamics. This method, which is also termed the dimensional-regularization prescription, was employed previously in \cite{cognola2013einstein} and has resulted in a significant amount of interest in the proposed 4D-EGB gravity
\cite{guo2020innermost,fernandes2020charged,wei2020extended,nasir2023influence}.

Oikonomou \emph{et al.} \cite{oikonomou2024einstein} looked at the EGB theories that were harmonious with GW170817 at the end of the inflationary epoch and during the reheating period. Their work focused on the calculations of the scalar coupling function during a set of different evolutionary epochs. Dialektopoulos \emph{et al.} \cite{dialektopoulos2022dynamical} investigated the dynamical system in the EGB theory. To evaluate the stability criteria of this system, they examined its key points. Compared to earlier studies in the literature, we identified new critical moments that could be crucial for comprehending the broader evolution of cosmology within gravitational models. Shahidi and Khosravi \cite{shahidi2022anisotropy} compared the theory with observational data and took into account the anisotropic cosmology. Through radial, linear perturbations, the stabilities of the thin shell wormhole model in 4D EGB gravity are examined by Zhang \emph{et al.} \cite{zhang2023traversable}. According to Israel junction requirements, these solutions are usually traversable and have a thin-shell throat. These stellar models were noticed to be stable when accompanied due to exotic matter as the charge increases in significance. Doneva \emph{et al.} \cite{doneva2021relativistic} performed numerical simulations to evaluate some exact analytical solutions of the relativistic ideal stellar structures in this theory. To demonstrate the impact of this gravity, we aim to understand the grounds under which irregularities appear in a homogeneous distribution of spherical geometries in D-dimensional modified gravity.

In astrophysics, gravitational collapse is an occurrence that usually happens when a giant star runs out of nuclear fuel at its core. New celestial objects like black holes or neutron stars might arise as a result of this collapse \cite{joshi2007gravitational}. A star's life is characterized by an equilibrium between the pressure exerted outward by nuclear fusion processes and the inward pull of gravity. When a star faces inefficiency of fuel then the gravitational pull of its gravity leads to the contraction of the stellar core. If its core mass is beyond a critical level (Chandrashekhar) limit \cite{thorsett1996gravitational}, the star could implode to the formation of stellar objects. In a collapsing structure, the energy density distribution is not necessarily homogeneous. Inhomogeneities may result from differences in the composition and dispersion of materials inside the collapsing region. The dynamical characteristics of the collapsing object may be influenced by these inhomogeneities, resulting in the emergence of objects with different densities.

Dwivedi and Joshi \cite{dwivedi1992cosmic} searched for naked singularities in the inhomogeneous collapse of gravity employing the Tolman-Bondi model. The collapse analysis in a compact spherical distribution was observed by Herrera \emph{et al.} \cite{herrera1998role}. Gravitational properties of dissipative spherical distribution are examined by Thirukkanesh and Maharaj
\cite{thirukkanesh2009radiating}. Herrera \emph{et al.} \cite{herrera2012cylindrically} proposed a general review of fluid configuration for the geodesic movement based on the structure scalar in cylindrically symmetric geometry. The dynamical characteristics of a charged radiating fluid are controlled by the tilted congruences as examined by Yousaf \cite{yousaf2021stable}. He looked at irregular density stability for Maxwell-Palatini $f(R)$ theory. Maurya \emph{et al.}
\cite{maurya2020buchdahl} investigated the equilibrium state of star development, with consequences for several cosmological and astronomical problems. They searched every possible Buchdahl solution in the theory under consideration and then contrasted them with GR. To examine the stability and survival of stars, they looked into effective factors including tangential $(P_{\bot})$ pressure, radial $(P_{r})$ pressure, anisotropy, and density. In the framework of general GR, Maurya \emph{et al.} \cite{maurya2018role} developed an anisotropic self-gravitational object with objects maintaining spherical geometry. They calculated some exact analytical models to understand the local-anisotropic pressure role by applying the Orlyanskii and Korkina metric potential. Kanti \emph{et al.} \cite{kanti2015gauss} asserted that higher curvature parameters are likely to be the most important in the study of the universe's higher-energy domain, to demonstrate the relevance of $f(\mathcal{G})$ elements in inflation. The occurrence of analytical and approximate solutions in the context of the various backgrounds was examined recently by  \cite{kanti2011wormholes,yousaf2016causes}.

Mena and  Tavakol \cite{mena1999evolution} conducted a comparative analysis within the framework of Lema\^{i}tre-Tolman and Szekeres irregular cosmological models. Sharma and Tikekar \cite{sharma2012space}
studied the anisotropic shear-free collapsing composition and found
that inhomogeneity significantly affects the various characteristics
of the collapsing fluid. Yousaf \emph{et al}. \cite{yousaf2023energy} examined a variety of fluid types, including isotropic, pressure-less, and anisotropic fluids, to judge the stability of a radiating spherical configuration. Using insights from the anisotropic composition of the radiating fluid under consideration, Yousaf \emph{et al}. \cite{yousaf2023sources} examined the inhomogeneity for the relativistic charged system. Astashenok et al. \cite{govender1998causal2} performed some numerical simulations to discover stable configurations of compact stars in GB gravity.

This research aims to investigate the reasons for the density inhomogeneity in the initially homogeneous spherically anisotropic distribution  \cite{herrera2024cracking,herrera2018new,herrera2020stability,herrera2011physical} within the framework of 4D-EGB gravity. The various sections are organized as follows: Section II comprises the fluid distribution in
the anisotropic spherically symmetric geometry, which is confined by $\Sigma$. This section further discusses the kinematics of spheres. The field equations accompanied by mass variational equations, Ellis equations, and dynamical equations for the proposed theory are discussed in Section III. Section IV is devoted to analyzing the causes of irregularities within various relativistic matter contents in $ D$-dimensional modified gravity. Lastly, we summarize our findings in Section V.

\section{Kinematical Quantities of Fluid Spheres}

We examine the collapse of a fluid distribution having a spherical
orientation, with a spherical surface $\Sigma$ serving as the
fluid's boundary. The field equations, which establish a connection
between spacetime's curvature and the distribution of both energy
and matter inside it, can be used to investigate the collapse of
such fluids. We also assume that the fluid has a local anisotropic pressure,
and we represent the interior metric using comoving coordinates
inside the surface $\Sigma$. In its generic form, the interior
metric is quantified by
\begin{align}\label{1}
ds^{2} =-C^2(t,r)dt^{2} +X^2(t,r)dr^{2} +Y^2(t,r){(d\theta^2
+\sin^2\theta d\phi^2)},
\end{align}
where it is assumed that $C,~ X,$ and $Y$ are positive. In contrast to the
Schwarzschild metric, which depicts the geometry outside the
surface, this metric reflects the inside geometry of the fluid. The
matter-energy distribution in spacetime can be expressed as the
matter-energy $T_{\varrho\varsigma}$, which is essential for
determining dynamical equations. It is given as
\begin{align}\label{b}
&T^{(m)}_{\varrho
\varsigma}=V_\varsigma
V_\varrho(\mu+P_{\bot})+q_\varrho
V_\varsigma-\chi_\varsigma\chi_\varrho(P_\bot -P_r)
+ V_\varrho
q_\varsigma+ \epsilon l_\varsigma
l_\varrho+P_{\bot}g_{\varrho\varsigma}.
\end{align}
The physical quantities $\mu,~P_{\bot},~P_{r},~q_{\varsigma},~
\epsilon,~V_{\varsigma},~\chi_{\varsigma}$ and~$l_{\varsigma}$ that
make up the expression have particular meanings in this particular
context. The fluid's pressure in the direction of motion is
symbolized by $P_{r}$. Here, $P_{\bot}$ and $\mu$ stand for the
pressure perpendicular to the motion and the energy density,
respectively. The components of four-velocity in the $\varsigma$
and $\varrho$ directions are denoted by the symbols $V_\varsigma$
and $V_\varrho$, respectively. The heat flux $q_{\varsigma}$
pointing in the direction of velocity is shown by the second term.
The anisotropic pressure is represented by the third term, where
$\chi_{\varsigma}$ denotes a radially directed unit vector. In the second last term $\epsilon$ and $l_{\varsigma}$
stand for null energy density and the associated four
vector, respectively. The tangential pressure, or the pressure in
the directions perpendicular to the motion is represented by the
last term, where the metric tensor, or $g_{\varrho\varsigma}$,
symbolizes the spacetime geometry. These quantities meet certain
requirements as
\begin{align}\label{4}
l^{\upsilon}l_{\upsilon}=0,~V_{\upsilon}\chi^{\upsilon}=0,~V_{\upsilon}l^{\upsilon}=-1,
\\\label{5}
\chi^{\upsilon}\chi_{\upsilon}=1,~V^{\upsilon}V_{\upsilon}=-1,~q_{\upsilon}V^{\upsilon}=0.
\end{align}
We can use the following equations to find the fluid's expansion and
four-acceleration.
\begin{align}\label{6}
\Theta =V^{\upsilon}~_{; \upsilon}~,~a_{\upsilon} =V^{\varsigma}
V_{\upsilon ; \varsigma}.
\end{align}
Furthermore, we may use the expression to find the fluid's shear
$\sigma_{\varsigma\upsilon}$
\begin{align}\label{7}
\sigma_{\varsigma\upsilon} = a_{(\varsigma V_\upsilon)}-
\frac{1}{3}\Theta h_{\varsigma\upsilon} +V_{(\varsigma ; \upsilon)},
\end{align}
where~$h_{\varsigma\upsilon} =V_{\upsilon}V_{\varsigma}
+g_{\varsigma\upsilon}$. Since we have assumed that the metric taken
into account in Eq. (\ref{1}) is comoving, then we have
\begin{align}\label{3}
l^{\upsilon} =C^{-1}\delta^{\upsilon}_{1} +
X^{-1}\delta^{\upsilon}_{2},~~\chi^{\upsilon}
=X^{-1}\delta^{\upsilon}_{2},~~V^{\upsilon}
=C^{-1}\delta^{\upsilon}_{1},~~ q^{\upsilon}
=qX^{-1}\delta^{\upsilon}_{2},
\end{align}
where $q^{\varsigma}$ is defined as a  function of $t$ and $r$
obeying $q^{\upsilon} = q \chi^{\upsilon}$. We define the non-zero component of
$a_{\upsilon}$ as
\begin{align}\label{8}
a^{2} =a_{\varsigma}a^{\varsigma}=
\bigg(\frac{C'^{2}}{X^{2}C^{2}}\bigg),~a_{2} =\frac{C'}{C}.
\end{align}
Here, $a^{\varsigma} = a \chi^{\varsigma}$, in which $a$ is a scalar. For expansion scalar, we have
\begin{align}\label{9}
\Theta=\frac{1}{C}\bigg(\frac{\dot{X}}{X} +\frac{2\dot{Y}}{Y}\bigg).
\end{align}
By using Eq. (\ref{3}) in Eq. (\ref{7}), we get
\begin{align}\label{10}
\sigma_{22} =\bigg(\frac{2}{3}~\sigma\bigg)X^{2} ,~\sigma_{33}
=\frac{\sigma_{44}}{\sin^{2}\theta} = -\bigg(\frac{1}{3}~\sigma
\bigg)Y^{2}.
\end{align}
Here, the scalar takes the form
\begin{align}\label{11}
\sigma_{\varsigma\upsilon}\sigma^{\varsigma\upsilon}=\frac{2}{3}\sigma^{2},~~
\sigma =\frac{1}{C}\bigg(\frac{\dot{X}}{X} -\frac{\dot{Y}}{Y}\bigg).
\end{align}

\section{Einstein Gauss-Bonnet Gravity in $D$-dimensional spacetime}

Now, we will examine the Einstein's gravity within a $D$-dimensional
spacetime, integrating a GB term and the cosmological constant
$\Lambda$. The generic action of the following form characterizes
this $D$-dimensional EGB gravity as
\begin{align}\label{a}
\mathcal{S}=\int\bigg[\mathcal{L}_{m}+\frac{\alpha}{D-2}R-\Lambda\bigg]\sqrt{-g}d^D
x,
\end{align}
where $g\equiv g^{\varsigma\varrho}g_{\varsigma\varrho}$ and  $\mathcal{L}_{m}$, is the matter Lagrangian density. The GB coupling constant $\frac{\alpha}{D-2}$ determines the weight of the GB term, which is proportional to the $R$ (scalar curvature). In 4D, this term is a topological invariant, but in higher dimensions, it plays a role in the dynamics. The energy density for the vacuum is symbolized by the cosmological constant or $\Lambda$. By varying
the above action field equations can be found as
\begin{align}\label{c}
\frac{2\alpha}{D-2}\bigg(R_{\varsigma\upsilon}-\frac{1}{2}Rg_{\varsigma\upsilon}\bigg)+\Lambda
g_{\varsigma\upsilon}=\kappa T_{\varsigma\upsilon}.
\end{align}
Applying Eqs. (\ref{b}), (\ref{3}), and (\ref{c}), we have
\begin{align}\label{16}
&[{\mu +\epsilon} +\Lambda]C^{2}
=\frac{2\alpha}{D-2}\bigg[\bigg[-\bigg(\frac{C}{X}\bigg)^{2}\bigg[\bigg(\frac{Y'}{Y}\bigg)^{2}-2\frac{X'Y'}{XY}-
\bigg(\frac{X}{Y}\bigg)^{2}+\frac{2Y''}{Y}\bigg]+\frac{\dot{Y}}{Y}\bigg(\frac{2\dot{X}}{X}+\frac{\dot{Y}}{Y}\bigg)\bigg],
\\\label{17} &[{P_{r} +\epsilon}
-\Lambda]X^{2}=\frac{2\alpha}{D-2}\bigg[\frac{Y'}{Y}\bigg(\frac{Y'}{Y}+\frac{2C'}{C}\bigg)-
\bigg(\frac{X}{Y}\bigg)^{2}-\bigg(\frac{X}{C}\bigg)^{2}\bigg[\frac{2\ddot{Y}}{Y}-\bigg(\frac{2\dot{C}}{C}
-\frac{\dot{Y}}{Y}\bigg)\frac{\dot{Y}}{Y}\bigg]\bigg], \\\label{18}
&[q+\epsilon]CX
=\frac{2\alpha}{D-2}\bigg[2\bigg(\frac{\dot{Y'}}{Y}-\frac{\dot{X}{Y'}}{XY}-\frac{\dot{Y}C'}{CY}\bigg)\bigg],
\\\nonumber &[P_\bot  -\Lambda]Y^{2}
=\frac{2\alpha}{D-2}\bigg[\bigg(\frac{Y}{X}\bigg)^2\bigg[\frac{Y''}{Y}+\frac{C''}{C}-\frac{C'X'}{CX}+\frac{Y'}{Y}\bigg(\frac{C'}{C}-
\frac{X'}{X}\bigg)\bigg]-\\\label{k2}&\bigg(\frac{Y}{C}\bigg)^2\bigg[\frac{\ddot{Y}}{Y}+\frac{\ddot{X}}{X}+\frac{\dot{X}\dot{Y}}{XY}
-\frac{\dot{C}}{C}\bigg(\frac{\dot{X}}{X}+\frac{\dot{Y}}{Y}\bigg)\bigg]\bigg].
\end{align}
\begin{figure}[H]
\centering
{{\includegraphics[height=2.0 in, width=3.0 in]{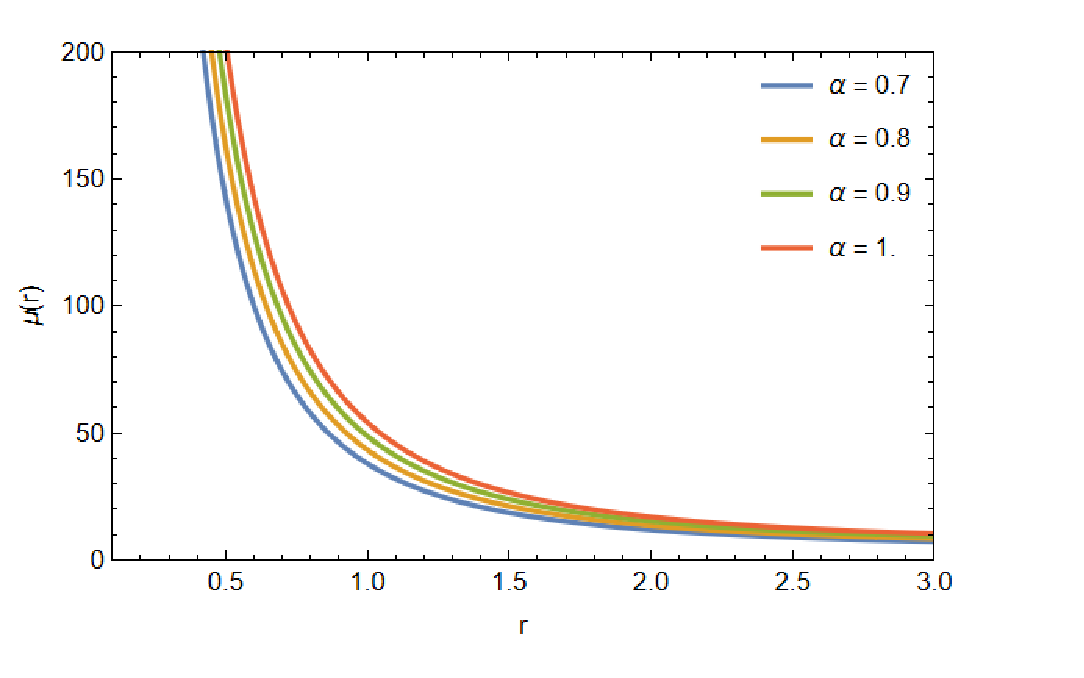}}}
\quad{{\includegraphics[height=2.0 in, width=3.0 in]{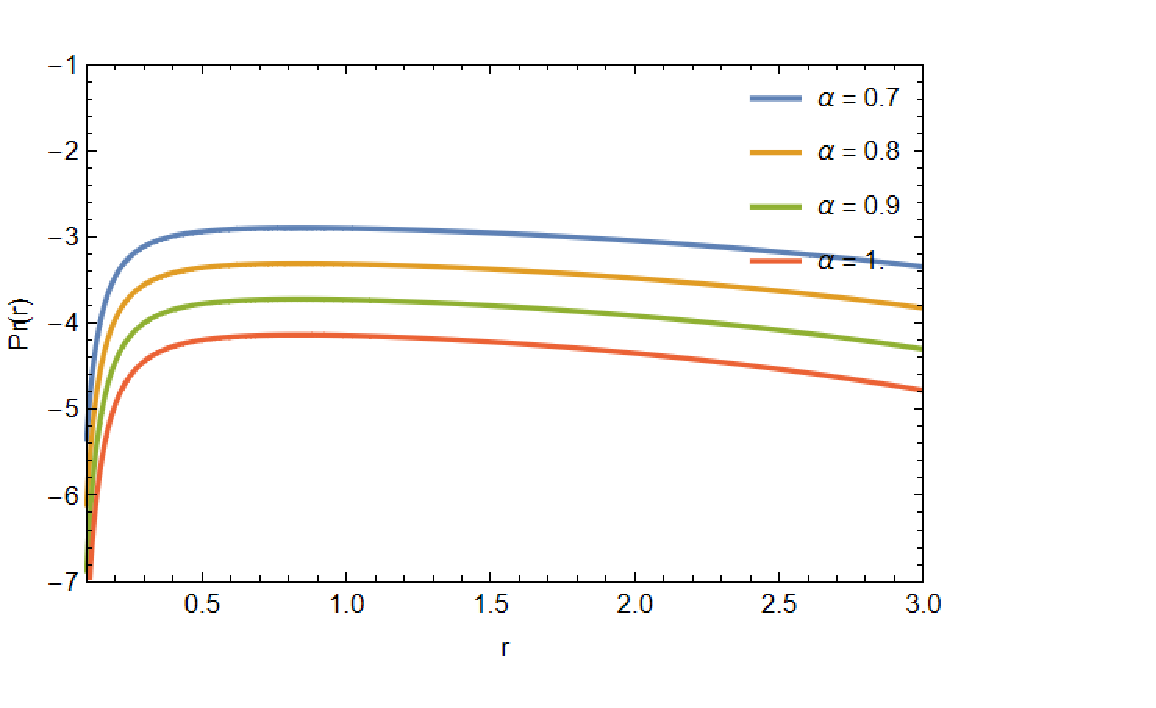}}}
\quad{{\includegraphics[height=2.0 in, width=3.0 in]{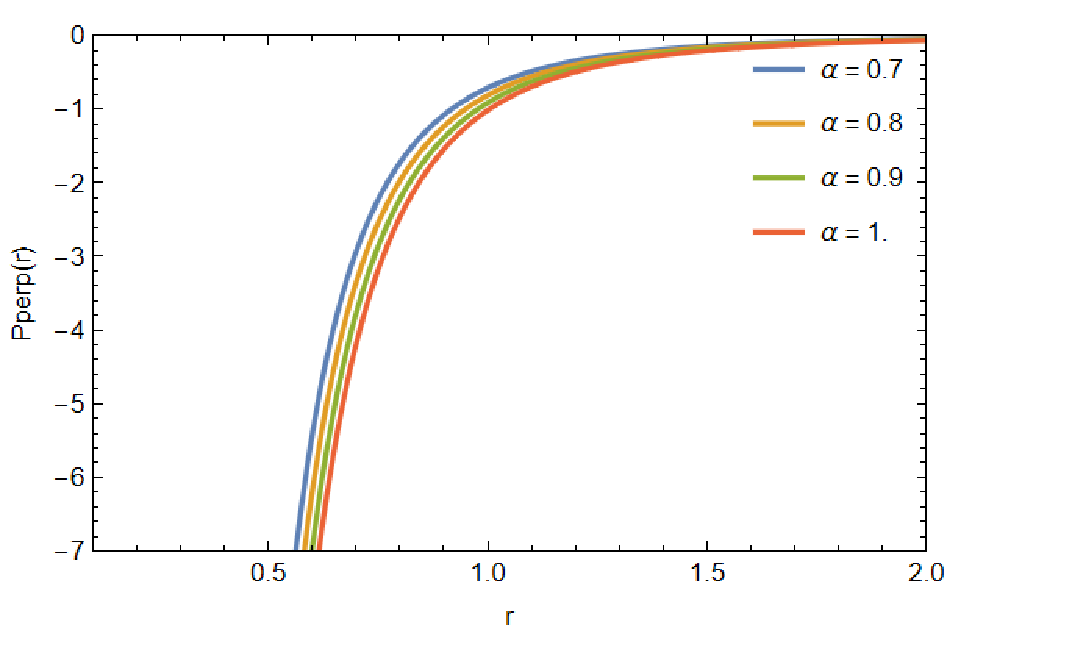}}}
\caption{Graph   of energy density $\mu$, the radial pressure $p_{r}$, and the tangential pressure $P_{\bot}$, as a function of $r$, for various values of the coupling constant, $\alpha$ .
}\label{1}
\end{figure}

The energy density $\mu$ in Fig. \textbf{\ref{1}} reduces as $r$ increases, indicating that the energy distribution is concentrated around small values of $r$ and diminishes as one advances outward.
Higher coupling constants result in a higher energy density across the region because, for a fixed $r$, $\mu$ increases as $\alpha$  grows.
$p_{r}$ is negative, suggesting that the force connected to the system may be attractive. $p_{r}$ becomes more negative as $\alpha$ increases, indicating that a stronger coupling results in a more noticeable radial pressure. Over the specified range of $r$, the tangential pressure likewise stays negative. $P_{\bot}$ becomes closer to zero as $r$ rises.
$P_{\bot}$, which exhibits a similar pattern to the radial pressure, is more negative for larger values of $\alpha$.

Thus, the findings demonstrate that a higher energy density and more negative radial and tangential pressures occur from raising the coupling constant $\alpha$.  This implies that gravitational impacts are amplified by increased coupling, which affects the stability and structure of the system.   All things considered, the physical properties of the system are greatly influenced by $\alpha$.

Misner and Sharp \cite{misner1964relativistic,misner1965spherical} provided a helpful formula to investigate the matter quantity for the spherical distribution as
\begin{align}\label{k1}
 &m = R^{232}_{3}\frac{Y}{2} =
\bigg(\frac{Y}{2} -\bigg(\frac{Y'}{X}\bigg)^{2}
+\bigg(\frac{\dot{Y}}{C}\bigg)^{2}\bigg).
\end{align}
The term $m=m(t, r)$ reflects the mass function, which is a measure of a compact fluid composition with radius $r$ and a center at any given time $t$ in spacetime. The derivative operators for the proper time and areal radius are provided as
\begin{align}\label{20}
D_{T} =\bigg(\frac{1}{C}\bigg)\frac{\partial}{\partial{t}},~~D_{Y}
=\bigg(\frac{1}{Y'}\bigg)\frac{\partial}{\partial{r}}.
\end{align}
As a variation in the areal radius over proper time, $D_{T}$ allows
to figure out the collapsing fluid velocity as
\begin{align}\label{k}
U =D_{T}Y.
\end{align}
This definition of $U$ is the rate of change of the areal radius as determined by an observer that moves with the fluid.  Equation (\ref{k1}) can be written after using Eq. (\ref{k}) as
\begin{align}\label{23}
{E} \equiv\frac{Y'}{X} =\bigg[1 -\frac{2m}{Y} +U^{2}
\bigg]^\frac{1}{2}.
\end{align}
The variational mass is calculated in terms of the relevant quantities by using Eq. (\ref{k1}) and the associated field equations as
\begin{align}\label{19}
&D_{T}m
=-Y^{2}\bigg[(\tilde{P_{r}}-\Lambda)U+E\tilde{q}\bigg]\frac{D-2}{2\alpha},
\\\label{20} & D_{Y}m=Y^{2}\bigg[(\tilde{\mu}+\Lambda)+\frac{\tilde{q}U}{E}\bigg]\frac{D-2}{2\alpha}.
\end{align}
Equation (\ref{20}) yields
\begin{align}\nonumber
&m =
\int^{r}_{0}\frac{D-2}{2\alpha}\bigg((\tilde{\mu}+\Lambda)+\frac{\tilde{q}U}{E}\bigg)Y^{2}Y'dr.
\end{align}
The above equation yields the specific relation between matter
variable and the factor symbolizing GB term. One can write
\begin{align}\label{21}
&\frac{3m}{C^{3}}
=\frac{D-2}{2\alpha}(\tilde{\mu}+\Lambda)-\frac{1}{Y^{3}}
\int^{r}_{0}\frac{D-2}{2\alpha}\bigg[D_{Y}\tilde{\mu}-3\frac{\tilde{q}U}{YE}\bigg]Y'Y^{3}dr,
\end{align}
where $\tilde{\mu} =\epsilon+\mu,~\tilde{q} =\epsilon+q$. The Vaidya spacetime is now taken into consideration as an external metric that characterizes the outgoing radiation in the form
\begin{equation}
ds^2 = \left[ - \left( 1 - \frac{2M(v)}{r} \right) dv^2 - 2 dv dr + r^2 (d\theta^2 + \sin^2\theta d\phi^2) \right].
\end{equation}
In this case, the retarded time and total mass are represented by $M(\nu)$ and $\nu$, respectively. Using the hypersurface $r = r_{\Sigma}=$ constant, we shall formulate the smooth matching between the charged Vaidya metric and the spherically symmetric metric. The continuity of the first and second fundamental forms over the hypersurface ${\Sigma}$, expressed as $(ds^2)_{\Sigma} = (ds^2_{-}) = (ds^2_{+})_{\Sigma}$, $[K_{\rho \lambda}] = K^{+}_{\rho \lambda} - K^{-}_{\rho \lambda} =0$, respectively, will be used to discuss the Darmois matching conditions, where ${\rho}, {\lambda} = 0, 2, 3$, and $K_{\rho \lambda}$ is the extrinsic curvatures, written as
\begin{equation}
K^\pm_{\rho \lambda} = -n^{\pm}_{\alpha}\left( \frac{\partial^{2} x^{\alpha}_{\pm}}{\partial {\eta}^{\rho} \partial {\eta}^{\lambda}}  + {\Gamma}^{\beta}_{\xi \varsigma} \frac{\partial^{2} x^{\xi}_{\pm}}{\partial {\eta}^{\rho}}  \frac{\partial^{2} x^{\varsigma}_{\pm}}{\partial {\eta}^{\lambda}} \right).
\end{equation}
In this case, $n^{\pm}$ is the outward unit normal over the hypersurface. The coordinates of the inner and exterior metric are denoted by $x^{\alpha}_{\pm}$, and ${\eta}^{\lambda}$ are the coordinates on the boundary. For the continuity of spacetime, we obtain
\begin{equation}
\frac{dv}{dt} \overset{\Sigma}{=} \frac{1}{C}, \quad N(t, r_\Sigma) = r_\Sigma, \quad \frac{dv}{dt} \overset{\Sigma}{=} \left( 1 - \frac{2M}{r}  + \frac{2dr}{dv} \right)^{-1/2}.
\end{equation}
The non-vanishing components of the extrinsic curvature are
\begin{align}\nonumber
&K^{-}_{00} = \left[ -\frac{C'}{C B} \right]_\Sigma, \quad
K^{-}_{22} = \frac{K^{-}_{33}}{\sin^{2}{\theta}} = \left[ \frac{N N'}{N} \right]_\Sigma, \quad
K^{+}_{00} = \left[ \left(\frac{d^{2}v}{d{\tau}^{2}}\right)^{-1}  \left( \frac{dv}{d{\tau}} \right)^{-1}  \frac{M}{r^{2}}\left( \frac{dv}{d{\tau}} \right)\right]_{\Sigma}, \\\label{4}& K^{+}_{22} =  \frac{K^{+}_{33}}{\sin^{2}{\theta}}= \left[ \left(\frac{dr}{d{\tau}}\right)r  +\frac{dr}{d{\tau}}\left(  1- \frac{M}{r}\right)\left( \frac{dv}{d{\tau}} \right)\right]_{\Sigma}
\end{align}
After smooth matching of $K^{-}_{22}$ with $K^{+}_{22}$, $K^{-}_{00}$, and $K^{+}_{00}$, we obtain:
\begin{equation}
{m(t,r)_\Sigma = M(v)_\Sigma, \quad P_r - \Lambda \overset{\Sigma}{=} q.}
\end{equation}
where on the hypersurface both the function of mass and the total mass $M(v)$ equalize.

The transverse and traceless part of the curvature is portrayed by the
Weyl tensor is considered as a specific component of the
$R_{\varsigma\sigma\zeta\upsilon}$. In GR, the Weyl tensor is a mathematical instrument that describes the shape of spacetime in the absence of direct interaction from matter or energy.  It aids in our comprehension of how gravity travels across space, particularly in void areas or distant from large objects.  The Weyl tensor describes how gravitational waves and tidal forces propagate, in contrast to the Ricci tensor, which informs us about the existence of matter and energy.  Studying energy density inhomogeneities, or unequal energy distributions in space, is one of its key applications. The Weyl tensor is used in cosmology to observe how gravity influences the formation and evolution of galaxies and other cosmic structures.  It is essential to comprehend the universe's large-scale structure because it clarifies why some regions contain more matter than others.
Its expression is given as
\begin{align}\label{k5}
&C^{\varsigma}_{\upsilon\varrho\beta}
=R^{\varsigma}_{\upsilon\varrho\beta}
-\frac{1}{2}R^{\varsigma}_{\varrho}g_{\upsilon\beta}
+\frac{1}{2}R_{\upsilon\varrho}\delta^{\varsigma}_{\beta}
-\frac{1}{2}R_{\upsilon\beta}\delta^{\varsigma}_{\varrho}
+\frac{1}{2}R^{\varsigma}_{\beta}g_{\upsilon\varrho}
+\frac{1}{6}R\bigg(g_{\upsilon\beta}\delta^{\varsigma}_{\varrho}
-g_{\upsilon\varrho}\delta^{\varsigma}_{\beta}\bigg).
\end{align}
The above tensor is known to be represented by the electric and magnetic components \cite{mcintosh1994electric}. The latter under consideration vanishes due to its symmetry, leaving just the electric component. The final
component leads to
\begin{align}\label{29}
E_{\varsigma\upsilon}
=C_{\varsigma\rho\upsilon\beta}V^{\rho}V^{\beta},
\end{align}
whose non-vanishing components are
\begin{align}\label{d1}
&3E_{22} =2X^{2}\mathcal{E},~~ 3E_{33} =-Y^{2}\mathcal{E},~~
\frac{E_{44}}{E_{33}} =~\sin^{2}\theta.
\end{align}
Here, the gravitational effects in the relativistic structure by
the tidal forces are characterized by the scalar $\mathcal{E}$, given by
\begin{align}\label{k6}
&\mathcal{E}=\frac{1}{2}\bigg(\frac{X'Y'}{X^{3}C}
-\frac{Y''}{YX^{2}} +\frac{\ddot{Y}}{Y}
-\frac{\dot{Y}^{2}}{Y^{2}}\bigg)
+\frac{1}{2}\bigg(\frac{\dot{Y}\dot{X}}{XY} -\frac{\ddot{X}}{X}
-\frac{1}{Y^{2}} +\frac{Y'^{2}}{X^{2}Y^{2}}\bigg).
\end{align}
The expression that relates the matter variables, the mass of collapsing distribution along with the Weyl scalar and the additional EGB term is derived through Eqs.(\ref{16}), (\ref{17}), (\ref{k2}) with
Eqs.(\ref{k1}) and (\ref{21}) as follows
\begin{align}\label{22}
& \frac{D-2}{2\alpha}[\tilde{\mu} -\Pi+\Lambda]-\frac{3m}{Y^{3}}
 =\mathcal{E},
\end{align}
where $\Pi =-(P_{\bot}-\tilde{P_{r}})$.  The interpretation of the
conservation laws connected to the field equations depend
on the contracted Bianchi identities. These identities are helpful,
when examining the dynamics of anisotropic spheres. The two
independent components of dynamical equations yield
 \begin{align}\label{k3}
&\tilde{\dot{\mu}} +\frac{2P_{\bot}\dot{Y}}{Y}
+\frac{\tilde{P_{r}}\dot{X}}{X}+\tilde{\mu}\bigg(\frac{2\dot{Y}}{Y}+\frac{\dot{X}}{X}\bigg)
+2\tilde{q}\frac{(CY)^{'}}{XY} +\frac{\tilde{q'}C}{X} =0,
\\\label{k4}&\tilde{\dot{q}}+\frac{2\tilde{q}\dot{Y}}{Y}+\bigg(\tilde{P_{r}} +\tilde{\mu}\bigg)\frac{C'}{X}
+\frac{C}{X}\tilde{P_{r}}' + 2\Pi\frac{Y'C}{XY}
 +\frac{2\tilde{q}\dot{X}}{X} =0.
\end{align}
The Ellis equations \cite{stoeger1992fluid} are those that define a
relationship of matter variables along with Weyl tensor and EGB
term. These equations, which also go by the name evolution
equations include two parts: one has the temporal derivatives, and
the other has the spatial derivatives of the provided variables. It
is impossible to disregard the importance of these equations in
figuring out if a density homogeneity survives over the
celestial object. Ellis' method
\cite{ellis2012relativistic,ellis2009republication} is thus used for
figuring out these equations holds fundamental importance in the
calculation of inhomogeneity. Thus, the expressions are
\begin{align}\label{23}
&\bigg(\mathcal{E}-\frac{D-2}{2\alpha}(\tilde{\mu}
-\Pi+\Lambda)\bigg)^{.}=\frac{3\dot{Y}}{Y}\bigg((\tilde{\mu}+P_{\bot})\frac{D-2}{2\alpha}-\mathcal{E}\bigg)
+\frac{D-2}{2\alpha}\bigg(\frac{3\tilde{q}CY'}{XY}\bigg),\\\label{24}&\bigg(\mathcal{E}-\frac{D-2}{2\alpha}(\tilde{\mu}
-\Pi+\Lambda)\bigg)^{'}=-\frac{3Y'}{Y}\bigg(\mathcal{E}+\frac{D-2}{2\alpha}
\Pi\bigg)-\frac{D-2}{2\alpha}\bigg(\frac{3\tilde{q}X\dot{Y}}{CY}\bigg).
\end{align}

\section{Analysis of the inhomogeneity for relativistic fluids with various properties}

In this section, we will examine the factors that are responsible
for producing irregularities in different types of fluids. By systematic formulations under the framework of the provided theory, the quantities that disturb the regularity of the spherical fluids are studied one by one. For this, derived expressions for the Ellis
equations will help us in a better way.

\subsection{Geodesic fluid}

First, we consider  non-radiating  fluid with $C= 1$ and $P_{\bot}=\tilde{q}=P_{r}= 0$. Hence, Eqs. (\ref{23}) and (\ref{24}) for the pressureless fluid sphere turns out to be
\begin{align}\label{25}
&\bigg[\mathcal{E}-\frac{D-2}{2\alpha}({\mu}
+\Lambda)\bigg]^{.}=\frac{3\dot{Y}}{Y}\bigg[{\mu}\frac{D-2}{2\alpha}-\mathcal{E}\bigg],
\\\label{26}&\bigg[\mathcal{E}-\frac{D-2}{2\alpha}({\mu}
+\Lambda)\bigg]^{'}=-\frac{3Y'}{Y}\mathcal{E}.
\end{align}
It is evident from Eq. (\ref{26}) that $\mathcal{E}=0$ if and only
if $\mu^{'}=0$. To have more insight into the model, we arrange Eq. (\ref{25}) by using Eqs. (\ref{k3}) and (\ref{11}) in it. Then, it takes the following form
\begin{align}\label{27}
\dot{\mathcal{E}}
+\mathcal{E}\frac{3\dot{Y}}{Y}+\frac{D-2}{2\alpha}\mu\sigma=0.
\end{align}
The above equation suggests that the effect of irregularity, which is determined by fluid characteristics as well as the density of structures, is disturbed by the curvature tensor and coupling constant for the EGB term. The solution of Eq. (\ref{27})
is found as
\begin{align}\label{28}
\mathcal{E} =- \frac{\int^{t}_{0} Y^{3}
[\mu\sigma]\frac{D-2}{2\alpha}}{Y^{3}}~dt.
\end{align}
Here, $\mathcal{E}(0)= 0$ is regarded as the integration function. This establishes the requirements for dust fluid homogeneity. This implies that a system can only be shear-free, with no possibilities for curvature parameters, and conformally flat with an adiabatic homogenous distribution.
\subsection{Isotropic Fluid}

We now examine an adiabatic $(\tilde{q}=0)$ spherical system with
pressure that is locally isotropic $(\Pi= 0)$. Subsequently, the Ellis
equations become
\begin{align}\label{29}
&\bigg[\mathcal{E}-\frac{D-2}{2\alpha}({\mu}+
\Lambda)\bigg]^{.}=\frac{3\dot{Y}}{Y}\bigg[({\mu}+P)\frac{D-2}{2\alpha}-\mathcal{E}\bigg],
\\\label{30}&\bigg[\mathcal{E}-\frac{D-2}{2\alpha}({\mu}
+\Lambda)\bigg]^{'}=-\frac{3Y'}{Y}\mathcal{E}.
\end{align}
It can be observed that Eq. (\ref{30}) matches Eq. (\ref{26}), indicating that $\mathcal{E}=0$ if and only if $\mu^{'}=0$. This describes that the vanishing of the Weyl scalar is encouraging the system's homogeneity. By following the same technique Eq. (\ref{29}) is rearranged as
\begin{align}\label{31}
\dot{\mathcal{E}}
+\mathcal{E}\frac{3\dot{Y}}{Y}+\frac{D-2}{2\alpha}(\mu+P)C\sigma=0,
\end{align}
whose solution admits
\begin{align}\label{32}
\mathcal{E} =- \frac{\int^{t}_{0} Y^{3}
[(\mu+P)C\sigma]\frac{D-2}{2\alpha}}{Y^{3}}~dt.
\end{align}
This suggests that as long as the configuration undergoes shear-free motion, its energy density will be uniform. Therefore, the stability of the homogeneous density established in this case is not affected by the isotropic pressure for the shear-free configuration. It is well-known that isotropic fluids and conformally flat spherically symmetric spacetime (without dissipation) are shear-free, but not vice versa \cite{herrera2011physical}. It is therefore possible to
presume that the fluid is shear-free without requiring conformal
flatness, and we get
\begin{align}\label{33}
\mathcal{E}
=\frac{D-2}{2\alpha}\int^{t}_{0}\frac{1}{Y^{3}}f(r)~\equiv F(r),
\end{align}
where $F(r)$ is an arbitrarily chosen function. The former expression suggests that even for shear-free cases, there are the responsible parameters for the irregularity,y, which are identified as curvature scalar and EGB coupling. Hence, if the fluid is expanding and initially has a relatively small value of $\mathcal{E}$ (non-zero), then it will remain during evolution and will play its role in inhomogeneity.

\subsection{Anisotropic $(\Pi \neq 0)$ Fluid}

In this particular case, the distribution of matter is anisotropic
but not dissipating $(\tilde{q}=0)$. Through their influence on the unequal distribution of matter and energy over space, anisotropic effects are essential in forming inhomogeneities. Different parts of space may be stretched or compressed differently as a result of this directional dependence, which could change the energy density.  The development and evolution of objects such as galaxies and galaxy clusters are influenced by cosmic shear, primordial fluctuations, and large-scale tidal pressures, all of which can result in anisotropies in cosmology. These anisotropic effects aid in developing density contrasts over time, intensifying inhomogeneities. Within this structure, Eqs.
(\ref{23}) and (\ref{24}) provide
\begin{align}\label{34}
&\bigg[\mathcal{E}-\frac{D-2}{2\alpha}({\mu}
-\Pi+\Lambda)\bigg]^{.}=\frac{3\dot{Y}}{Y}\bigg[({\mu}+P_{\bot})\frac{D-2}{2\alpha}-\mathcal{E}\bigg],
\\\label{35}&\bigg[\mathcal{E}-\frac{D-2}{2\alpha}({\mu}
-\Pi+\Lambda)\bigg]^{'}=-\frac{3Y'}{Y}\bigg[\mathcal{E}+\frac{D-2}{2\alpha}
\Pi\bigg].
\end{align}
The first thing to note about the latter is that, in contrast to the above cases, the quantity $(\Pi +\mathcal{E})$, rather than only the Weyl tensor and the EGB term, is now accountable for the survival of density inhomogeneity. Thus, the regular distribution can be achieved by taking the vanishing of the accountable factor along with additional curvature terms. Now to introduce the kinematical behavior, we use Eqs. (\ref{k3}) and
(\ref{11}) in Eq. (\ref{34}), which yield
\begin{align}\label{36}
&\bigg(
\frac{D-2}{2\alpha}\Pi+3\mathcal{E}\bigg)\frac{\dot{Y}}{Y}+\bigg(
\Pi +\mathcal{E}\bigg)^.
=-\frac{D-2}{2\alpha}(\mu+P_{r})C\sigma+\frac{D-2}{2\alpha}\bigg(\dot{\Pi}\bigg(\frac{2\alpha}{D-2}-1\bigg)\bigg).
\end{align}
We define a tensor $X_{\varsigma\varrho}$
\cite{herrera2012cylindrically} as
\begin{align}\label{53}
&X_{\varsigma\upsilon}=^*R^*_{\varsigma\zeta\upsilon\nu}V^{\zeta}V^{\nu}=\frac{1}{2}\upsilon^{\varrho\delta}_{\varsigma\zeta}~R^*_{\varrho\delta\upsilon\nu}V^{\zeta}V^{\nu},
\end{align}
where $R^{*}_{\varsigma\zeta\upsilon\nu}$ = $\frac {1}{2}
\eta_{\epsilon\delta\upsilon\nu}R^{\omega\delta}_{\varsigma\zeta}$
and $\eta_{\omega\delta\upsilon\nu}$ stands for Levi-Civita tensor.
\begin{align}\label{54}
X_{\varsigma\upsilon} =\frac{1}{3}X_{T}h_{\varsigma\upsilon}
+X_{TF}\bigg(\chi_{\varsigma}\chi_{\upsilon}
-\frac{1}{3}h_{\varsigma\upsilon}\bigg).
\end{align}
The latter expression consists of two parts, one is trace-free and the other one is a trace part of Eq. (\ref{53}). After taking Eqs. (\ref{54}), (\ref{k5}), and (\ref{k6}) with field equations, we have the following
\begin{align}\label{55}
X_{TF} =-(\mathcal{E}+ \Pi).
\end{align}
It is well known that $X_{TF}$ \cite{herrera2010lemaitre} is a structural scalar which incorporates the factor $(\mathcal{E} + \Pi)$. Thus the definition of structure scalar $(X_{TF})$ can be used to address the density inhomogeneity when dissipation is absent.  Consequently, the evolution equation (\ref{36}) reads
\begin{align}\label{37}
&{\dot{X_{TF}}} +X_{TF}\frac{3\dot{Y}}{Y}
 = \frac{D-2}{2\alpha}(P_{r}+\mu)C\sigma-\frac{D-2}{2\alpha}
 \bigg[\dot{\Pi}\bigg(\frac{2\alpha}{D-2}-1\bigg)-\frac{\Pi\dot{Y}}{Y}
 +\frac{2\alpha}{D-2}\frac{3\Pi\dot{Y}}{Y}\bigg],
\end{align}
whose generic solution can be identified as
\begin{align}\label{39}
&X_{TF} =\frac{
\int^{t}_{0}\frac{D-2}{2\alpha}[(\mu+P_{r})C\sigma]Y^{3}}{Y^{3}}~dt-\frac{\int^{t}_{0}[\dot{\Pi}(\frac{2\alpha}{D-2}-1)-\frac{\Pi\dot{Y}}{Y}
 +\frac{2\alpha}{D-2}\frac{3\Pi\dot{Y}}{Y}]Y^{3}}{Y^{3}}~dt.
\end{align}
It suggests that one of the structure scalars is the quantity controlling the stability of inhomogeneous density. Thus, in addition to the additional degree of freedom provided by 4D-EGB theory, the initial homogenous state is influenced by the shear scalar and anisotropy of the fluid.

Using the spherically symmetric metric on the star object, SAX J1808.4-3658, we set up a relationship with the $X_{TF}$ and the theoretical concepts of irregularity and true observable signals. Neutron stars are produced by the remnants of supernova explosions and are incredibly compact objects.  They provide a useful laboratory for investigating the relationship between curvature and inhomogeneities because of their powerful gravitational fields and extreme densities. We investigate the irregularity of spherically symmetric collapsing anisotropic composition within 4D-EGB gravity. We relate our ideas to investigate the applicability and validity of our theoretical framework on the well-studied and observationally obvious compact object SAX J1808.4-3658. Observational data of the star candidate SAX J1808.4-3658 from \cite{yousaf2018existence} and the Krori-Braua ansatz \cite{krori1975singularity} were employed to look into the role of the inhomogeneity factor for an anisotropic adiabatic fluid.
The behavior of this factor is depicted in Fig. \textbf{\ref{1f}} in the Appendix.
%Figure 1 provides a thorough analysis of that factor.
The center part of a star suffers higher temperatures and pressures during its collapse, which results in larger concentrations of energy. Because the material at the center of the fluid is compressed more by gravity than at the surface, inhomogeneity in temperature and density becomes more noticeable as the fluid collapses. The formation of objects like neutron stars and black holes depends on the presence of inhomogeneity, which might take the form of increased kinetic energy and density gradients.
It is evident from Fig. \textbf{\ref{1f}} that the inhomogeneity factor plays a significant role in maintaining the corresponding system at the center of a sphere. As the observer moves away from the center
towards the boundary surface, the inhomogeneity term decreases its role gradually until it approaches zero in the presence of 4D-EGB gravity. Therefore, we deduce that the gravitational force is weaker near the star's surface than it is in the center, which causes the inhomogeneity to diminish. The outer layers may remain more consistent because they are not compressed as much. For an understanding of how stellar structures evolve during collapse, this contrast between the surface and the center is essential.

\subsection{Radiating Pressureless Fluid}

Lastly, we will examine the scenario of dissipative $(\tilde{q}\neq
0)$ geodesic $(C=1)$ dust to highlight the role of dissipation in
the development of irregularity. Note that, as it is evident from
Eq. (\ref{k4}), the dust condition in the dissipative case does not
entail that the fluid is geodesic. As a result, the geodesic
condition is assumed here for convenience as well as to separate the
effects caused solely by dissipative occurrences. It is important to
note that a lot of effort has been put into determining exact
solutions that describe dust geodesic forms
\cite{herrera2010lemaitre,govender1998causal,thirukkanesh2010mixed}.
After using the assumed conditions Eqs. (\ref{23}) and (\ref{24})
read
\begin{align}\label{41}
&\bigg[\mathcal{E}-\frac{D-2}{2\alpha}(\tilde{\mu}
+\Lambda)\bigg]^{.}=\frac{3\dot{Y}}{Y}\bigg[\tilde{\mu}\frac{D-2}{2\alpha}-\mathcal{E}\bigg]
+\frac{D-2}{2\alpha}\bigg[\frac{3\tilde{q}Y'}{XY}\bigg],\\\label{k7}&\bigg[\mathcal{E}-\frac{D-2}{2\alpha}(\tilde{\mu}
+\Lambda)\bigg]^{'}=-\mathcal{E}
\frac{3Y'}{Y}-\frac{D-2}{2\alpha}\bigg[\frac{3\tilde{q}X\dot{Y}}{Y}\bigg].
\end{align}
Equation (\ref{k7}) gives
\begin{align}\label{42}
\Psi \equiv~ \mathcal{E} +
\frac{\int^{r}_{0}3[\dot{Y}\tilde{q}X{Y^{2}}]\frac{D-2}{2\alpha}}{Y^{3}}~dr.
\end{align}
It demonstrates that $\Psi$ is now the factor in charge of
controlling inhomogeneity. It is easy to determine from
Eq. (\ref{k7}) that $\tilde{\mu'} = 0$  if and only if $\Psi = 0$.
Then, using Eqs. (\ref{k3}) and (\ref{11}), an  evolution expression
for $\Psi = 0$ may be found from Eq. (\ref{41}) as
\begin{align}\nonumber
\Omega
=3\int^{r}_{0}\bigg[\tilde{q}X\dot{Y}Y^{2}\bigg]\frac{D-2}{2\alpha}~dr.
\end{align}
 Equation (\ref{41}) admits
\begin{align}\label{k8}
 \frac{\dot{\Omega}}{Y^{3}}+\dot{\Psi} +\frac{3\dot{Y}\Psi}{Y} =
\frac{D-2}{2\alpha}\bigg[\tilde{\mu}\sigma+\frac{\tilde{q'}}{X}+\frac{\tilde{q}{Y'}}{XY}\bigg].
\end{align}
The generic solution of Eq. (\ref{k8}) is
\begin{align}\label{44}
\Psi = \frac{\int^{t}_{0}[\dot{\Omega}+
\frac{D-2}{2\alpha}[\tilde{\mu}\sigma+\frac{\tilde{q'}}{X}+\frac{\tilde{q}{Y'}}{XY}]]Y^{3}}{Y^{3}}~dt,
\end{align}
where it is evident how many variables have influenced the evolution
of $\Psi$. We may further modify the formulation above by observing
that in the case under investigation Eq. (\ref{k4}) may be formally
integrated to produce,
\begin{align}\label{45}
\tilde{q} =\frac{\phi(r)}{Y^{2}X^{2}},~~~~~~~~~~
\end{align}
where  $\phi(r)$ is an arbitrarily chosen, admiting $\phi(0) = 0$. It is evident from Eq. (\ref{44}) that the emergence of density inhomogeneities from a homogeneous structure is reliant upon distinct factors that are fluid shear and radiating terms under a 4D-EGB framework. Let us now examine the \textit{shear-free} case to better isolate the influence of these latter terms. It is easy to see that we can set $Y= Xr$ if the shear is supposed to vanish. Then, using Eq. (\ref{45}), we have from Eq. (\ref{44})
\begin{align}\label{43}
&\Psi
=\frac{\int^{t}_{0}\frac{D-2}{2\alpha}[\int^{r}_{0}\phi(r)\dot{\Theta}dr
+\frac{\phi^{'}(r)r^{3}}{Y^{2}}+\frac{\phi(r)r^{2}}{Y^{2}}(2-\frac{3Y'r}{Y})]~dt
}{Y^{3}}.
\end{align}
The factor governing irregularity has a certain association with fluid variables, specifically, scalar expansion $(\Theta)$, heat flow, and the structural dynamics of the distribution, which highlights the significance of fluid variables.

\section{Conclusions}

This work aims to analyze the role of 4D-EGB theory on the causes of energy-density irregularity of self-gravitating fluids. We also explain how those causes evolved from a homogeneous matter distribution. We noticed that the variation in energy density with 4D-EGB theory can have a noteworthy impact on a star's evolution, influencing several aspects like as energy dissipation, pressure distribution, heat distribution, and collapse behavior. The stability, life cycle, and ultimate fate of the star may be impacted by these effects. In the previous work, Herrera et al. \cite{herrera2011physical} analyzed the fate of inhomogeneity in the background GR. They did not invoke the correction terms emerging from such a kind of modified gravity. The present work is based on a theory that is obtained by adjusting the coupling associated with the GB term and $D$ dimensions for the modification in the field equations. The GB coupling constant $\frac{\alpha}{D-2}$ determines
the weight of the GB term, which is proportional to the $R$ (scalar curvature) incorporated in the field equations to retrieve the role of the local dynamics. To demonstrate the impact of this gravity, this work observes the irregularity behavior of
spherically symmetric distribution in the 4D-EGB framework. We have considered the spherically symmetric configuration with the geometry occupied by anisotropic fluid. The modified version of the field equations and the associated dynamical expressions have been
computed. The Misner-Sharp scheme is used to find the mass
function for our spherical distribution, and the Weyl tensor is
examined in this regard. It has been determined that the Weyl tensor comprises component tensors, such as magnetic and electric components.  The two differential equations that result from the
explicit representation of the curvature tensor with matter contents
and the mass function will be mandatory in our investigation.
Ellis's method, which he employed in his work, is applied to solve
these problems. After considering the following particulars under
the non-radiating and radiating structures, we can reach
the desired results.

For non-radiating geodesic dust, Eqs. (\ref{25}) and (\ref{26}) reveal the fact that the homogeneity of the distribution is influenced by the conformal tensor and coupling constant for the EGB term. From Eq. (\ref{28}), it is clear that the Weyl scalar is directly related to energy density, shear scalar, and the additional factor $\frac{2\alpha}{D-2}$, which incorporates the role of GB terms. Equation (\ref{28}) establishes the requirements for the dust fluid homogeneity. According to this, a system can only have a homogeneous configuration if it is conformally flat, and flatness requires the diminishing of shear scalar and parameters for curvature that are inducing irregularity in the configuration. This result is well-consistent with observational outcomes of gravitational collapse \cite{joshi2007gravitational} and supports the findings of \cite{herrera2011physical,yousaf2016causes}.

Then, we consider in non-adiabatic spherical system with isotropic pressure. It has been observed through Eq. (\ref{30}) that $\mathcal{E}=0$ if and only if $\mu^{'}=0$. This indicates that the vanishing of the Weyl scalar is encouraging the system homogeneity even in EGB theory. Consequently, Eq. (\ref{33}) suggests that even for shear-free cases, the responsible parameters for the irregularity are identified as curvature scalar and GB coupling parameters.

The irregularity in an anisotropic non-adiabatic system is expressed and then studied in detail. It is noticed that a term containing the impact of pressure anisotropy is involved in the production of irregularities. This factor is known as structure scalars, $X_{TF}$, and is stated in Eq. (\ref{37}). It has been noticed from Eq. (\ref{39}) that this structure scalar encapsulates EGB degree of freedom, shear scalar, and anisotropy of the fluid. The shear-free evolution gives more impact to the differences between the pressure components and EGB correction parameters in the maintenance of homogeneous spherically symmetric objects. Lastly, for radiating dust, the quantity $\Psi$ is investigated and found to be the cause of the formation of inhomogeneous energy density. From Eq. (\ref{44}), we have observed that this  $\Psi$ is reliant upon distinct factors, which are fluid shear, radiating terms, and EGB corrections.

Consequently, the energy density inhomogeneity of celestial self-gravitating structures is strongly affected by 4D-EGB gravity theories, which in turn affects the structure, stability, and complexity of these objects in alternative gravitational frameworks. The provided scenarios demonstrate that the regularity characteristics of the fluid are influenced by extra variables arising from the Gauss-Bonnet theory.

The physical reasons for energy-density homogenization and how it affects the stability of celestial bodies should be explored in more detail in future research. Knowing these systems can help us understand how stars behave and evolve. Exciting opportunities for further exploration exist in energy density inhomogeneity when studied in the context of several modified theories of gravity. Further investigation into the modified Gauss-Bonnet theory, like 4D-EGB gravity as a gravitational substitute for dark energy, may be undertaken. Investigating how this hypothesis affects energy density inhomogeneity and what it means for celestial bodies can lead to new insights into dark energy and gravity interactions.

Using the perturbative approach, Nashed and Sridakis \cite{nashed2022stability} studied the thermodynamics and motion stability in the case of spherically symmetric solutions in $f(T)$ gravity. They extracted charged black hole solutions for two charge profiles, namely with and without a perturbative correction in the charge distribution, taking into account minor departures from GR. They computed the energy and mass of the solutions, analyzed their asymptotic behavior, and extracted different torsional and curvature invariants. They found that the heat capacity is always positive for bigger deviations from GR. This indicates that $f(T)$ modifications enhance thermodynamic stability, whereas other classes of modified gravity do not.

Through the application of the Tolman-Finch-Skea metric and a particular anisotropy that is not directly dependent on it, as well as the smooth matching of the inner anisotropic solution to the Schwarzschild exterior one, Nashed and Saridakis \cite{nashed2023new} were able to extract new classes of anisotropic solutions within the framework of mimetic gravity. The data from the $4U 1608-52$ pulsar was then used to create a transparent image. They examined the radial and tangential pressures, as well as the energy density profile, and demonstrated that they are all positive and decrease toward the star center. However, our results demonstrate, the energy density is positive and decreases outward, which is consistent with the predicted physical behavior. But in contrast to the claim that radial and tangential pressure are positive, they both are negative throughout. Both pressures begin at a very negative value and progressively rise toward zero as $r$ increases, rather than falling toward the center.

We study the dynamical irregularity of the locally anisotropic spherical fluids in the context of Einstein-Gauss-Bonnet theory, where $\frac{\alpha}{D-2}$ determines the weight of the GB terms. We use the limit where the Gauss-Bonnet coupling constant  ${\alpha}\rightarrow0$ to examine the system behavior in the context of GR.  By doing this, the field equations are reduced to those of conventional Einstein gravity, essentially eliminating the higher-order curvature adjustments. The system is solely governed by the Einstein tensor $G_{\mu \nu}$, which is derived from the energy-momentum tensor of the anisotropic fluid. The extra contributions from the GB term disappear.

%%%%%%%%%%%%%%%%%%%%%%%%
%%%  Acknowledgments
%%%%%%%%%%%%%%%%%%%%%%%%
%\section*{Acknowledgments}

\section*{Acknowledgement}

The work of KB was partially supported by the JSPS KAKENHI Grant Numbers 21K03547 and 24KF0100. The work of MA was supported by the Researchers Supporting Project number: RSP2025R411, King Saud University, Riyadh, Saudi Arabia.

\section*{Conflict of Interest}

The authors declare no conflict of interest.

\section*{Data Availability Statement}

This manuscript has no associated data or the data will not be deposited. [Authors comment: This manuscript contains no associated data.]

\section*{Appendix}

\begin{figure}[H]
\centering
{{\includegraphics[height=3.0 in, width=5.0 in]{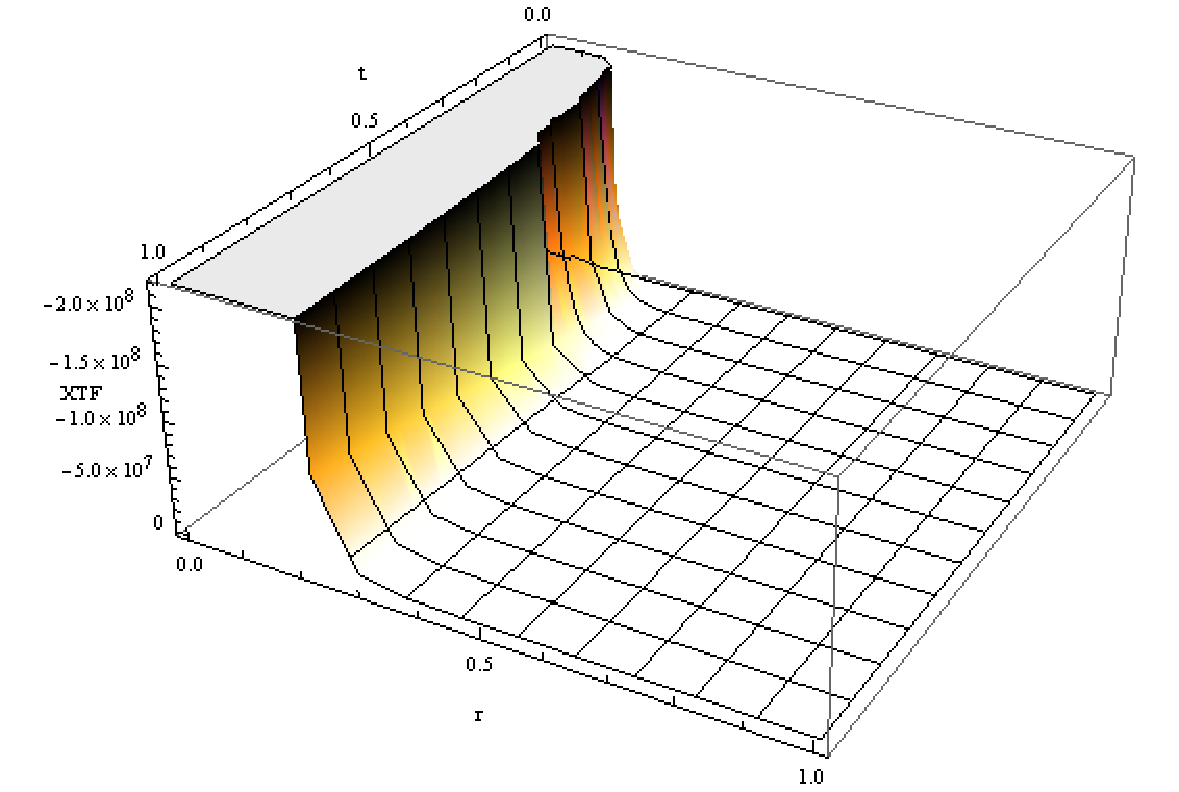}}}
\caption{Diagrammatic scheme of the factor of inhomogeneity $X_{TF}$ versus coordinates $r$ and $t$.}\label{1f}
\end{figure}

In this Appendix, we demonstrate the degree of variation in the inhomogeneity factor, which describes the change or fluctuation of various characteristics of the star like the star's density or other. In our study, it is noted that the inhomogeneity factor is high in the core due to the sharp variations in pressure, density, and EGB terms. It is also noticed through Fig. \textbf{\ref{1f}}  that the inhomogeneity factor decreases as we move outward because the gradients even out and the star is more uniform. It implies that as one moves outward from the core toward the surface, the star gets more homogeneous, or uniform.

\vspace{0.5cm}

\end{document}